\documentclass[preprintnumbers]{revtex4}

\usepackage{graphicx}
\usepackage{epsf}
\usepackage{amsmath,amssymb}
\usepackage{bm}
\usepackage{color}
\usepackage{textcomp}

\begin{document}
\preprint{KUNS-2527}

\title{Proton radii of Be, B, and C isotopes}
\author{Yoshiko Kanada-En'yo}
\affiliation{Department of Physics, Kyoto University, Kyoto 606-8502, Japan}

\begin{abstract}
We investigate the neutron number $(N)$ dependence of root mean square 
radii of point proton distribution (proton radii) of Be, B, and C isotopes with the theoretical method of variation after spin-parity projection
in the framework of antisymmetrized molecular dynamics (AMD). The proton radii 
in Be and B isotopes changes rapidly as $N$ increases, reflecting 
the cluster structure change along the isotope chains, 
whereas, those in C isotopes show a weak $N$ dependence because of the stable 
proton structure in nuclei with $Z=6$. 
In neutron-rich Be and B isotopes, the 
proton radii are remarkably increased by the enhancement of the two-center cluster structure
in the prolately deformed neutron structure. 
We compare the $N$ dependence of the calculated 
proton radii with the experimental ones reduced from the 
charge radii determined by isotope shift 
and those deduced from the charge changing interaction cross section. It is found that 
the $N$ dependence of proton radii can be a probe to clarify enhancement and weakening of 
cluster structures.

\end{abstract}
\maketitle

\section{Introduction}
In light unstable nuclei, 
various exotic structures such as the magic number breaking, 
new cluster structures, and the neutron halo structure, have been discovered.
In a series of Be isotopes, it has been revealed that 
the structure changes rapidly with the increase of the neutron number $N$ and the cluster structure 
develops in neutron-rich Be isotopes as discussed in 
many theoretical and experimental studies
\cite{Oertzen-rev,AMDsupp,KanadaEn'yo:2012bj,SEYA,OERTZENa,OERTZENb,KanadaEnyo:1995tb,ARAI,Dote:1997zz,ENYObe10,ITAGAKIa,ITAGAKIb,OGAWA,Descouvemont02,ENYObe11,KanadaEn'yo:2003ue,Ito:2003px,Ito:2005yy,FREER,SAITO04,Curtis:2004wr,Millin05,Freer:2006zz,Bohlen:2007qx,Curtis:2009zz,Yang14}.
The cluster structure in the ground states of $^{11}$Be and $^{12}$Be
is considered to play an important role in 
the vanishing of the neutron magic number $N=8$. 
For $^{11}$Be, the breaking of $N=8$ shell has been known experimentally from the 
abnormal spin-parity $1/2^+$, and for $^{12}$Be, it
has been suggested by slow $\beta$ decay \cite{Suzuki:1997zza} and 
more directly evidenced by the intruder configuration observed in 1$n$-knockout reactions \cite{Navin:2000zz,Pain:2005xw}
as well as other experiments \cite{iwasaki00,iwasaki00b,shimoura03}.
These nuclei have the largely deformed ground states with intruder neutron configurations
having more remarkable cluster structure than the neighboring isotope, $^{10}$Be. 

Also in neutron-rich B isotopes, the enhancement of cluster structures has been theoretically 
predicted \cite{KanadaEnyo:1995ir}, 
whereas, in neutron-rich C isotopes, no cluster structure is predicted to develop 
at least in the ground states \cite{AMDsupp,thiamova2004,KanadaEn'yo:2004bi}.
These facts indicate that the development of cluster structure strongly depends on 
proton and neutron numbers of the system. 
A problem to be solved is how one can experimentally observe the structure change along the isotope chain, i.e., 
the enhancement and weakening of the cluster structure with the increase of the neutron number $N$.
Since the enhanced cluster structure in neutron-rich nuclei enlarges 
the deformation and spatial extent of proton density, 
the change of the cluster structure may affect such observables 
as electric quadrupole moments and charge radii. 
The former is not necessarily a direct information of proton structure because 
it  is sensitive not only to the proton distribution but also to 
the neutron configuration through the angular momentum coupling. Moreover, 
it gives no information for the $J^\pi=0^+$ ground states of even-even nuclei, in which 
the quadrupole moment is trivially zero.
The latter, the charge radius, is usually not sensitive to the neutron configuration and it reflects 
more directly the proton density, at least for the radial extent, and therefore, 
the $N$ dependence of 
the charge radius can be a probe to clarify the change of the cluster structure.

Recently, root mean square (rms) charge radii of neutron-rich Be isotopes have been precisely measured by 
means of isotope shift. In the systematics of charge radii in Be isotopes, 
the large charge radii of $^{11}$Be and $^{12}$Be, which have been recently measured,
can be understood by the remarkable cluster structure in the deformed ground states of 
$^{11}$Be and $^{12}$Be \cite{Nortershauser:2008vp,Krieger:2012jx}. 
For neutron-rich B and C isotopes, change radii have yet to be measured except for 
$^{14}$C near the stability line. 
Instead of isotope shift measurement, recently, a new experimental approach to determine 
rms radii of point-proton density (proton radii) 
by the charge changing interaction cross section has been proposed and applied to B and C isotopes 
 \cite{Yamaguchi:2011zz,estrade14}.
 
Our aim here is to clarify how the structure change with the $N$ increase 
is reflected in proton radii.
For this aim, we investigate the $N$ dependence of proton radii in the isotope 
chains of Be, B, and C and the 
influence of change of cluster structures and intrinsic deformations on proton radii. 
We try to answer the question whether the $N$ dependence proton radii can be 
a probe for the cluster structure in neutron-rich nuclei. 

In this study, we calculate the ground states of Be, B, and C isotopes with the method of antisymmetrized molecular dynamics (AMD) \cite{AMDsupp}.
The method has been proven to be a useful approach to describe structures, in particular, cluster structures, in light 
neutron-rich nuclei. Systematic studies with the simple version of AMD have predicted that structures of 
Be and B isotopes change rapidly with the increase of the neutron number \cite{AMDsupp,KanadaEnyo:1995tb,KanadaEnyo:1995ir}. Advanced studies with the  
variation after spin and parity projections (VAP) in the AMD framework have described the breaking of $N=8$ magicity
in neutron-rich Be \cite{ENYObe11,KanadaEn'yo:2003ue}. 
The latter method (the AMD+VAP)
describes better the details of structures in ground and excited states than the former method
(the simple AMD), in which the variation is performed before the spin projection. 
In the present study, we apply the AMD+VAP to Be, B, and C isotopes, and discuss the 
structure change focusing on the $N$ dependence of the proton radius in each series of isotopes.

The paper is organized as follows.  
We describe the framework of the AMD+VAP
in Section \ref{sec:formulation}, and show the results of Be, B, and C isotopes  
in Section \ref{sec:results}. 
Section \ref{sec:discussion} discusses the structure change with the $N$ increase
and its influence on proton radii. 
The paper concludes with a summary in Section \ref{sec:summary}.

\section{Formulation of AMD+VAP}\label{sec:formulation}
We describe Be, B, and C isotopes with AMD wave functions by applying 
the VAP method. For the detailed formulation of the AMD+VAP, please refer to
Refs.~\cite{ENYObe10,ENYObe11,KanadaEn'yo:2003ue}.
The method is basically the same as that used in those previous studies.
A difference in the present calculation 
from Refs.~\cite{ENYObe10,ENYObe11,KanadaEn'yo:2003ue} is that
we do not adopt an artificial barrier potential, which has been used in previous studies
to describe highly excited resonance states. 

\subsection{AMD wave functions}

An AMD wave function is given by a Slater determinant of Gaussian wave packets;
\begin{equation}
 \Phi_{\rm AMD}({\bf Z}) = \frac{1}{\sqrt{A!}} {\cal{A}} \{
  \varphi_1,\varphi_2,...,\varphi_A \},
\end{equation}
where  ${\cal{A}}$ is the antisymmetrizer, and the $i$th single-particle wave function is written by a product of
spatial($\phi_i$), intrinsic spin($\chi_i$) and isospin($\tau_i$) 
wave functions as,
\begin{eqnarray}
 \varphi_i&=& \phi_{{\bf X}_i}\chi_i\tau_i,\\
 \phi_{{\bf X}_i}({\bf r}_j) & = &  \left(\frac{2\nu}{\pi}\right)^{4/3}
\exp\bigl\{-\nu({\bf r}_j-\frac{{\bf X}_i}{\sqrt{\nu}})^2\bigr\},
\label{eq:spatial}\\
 \chi_i &=& (\frac{1}{2}+\xi_i)\chi_{\uparrow}
 + (\frac{1}{2}-\xi_i)\chi_{\downarrow}.
\end{eqnarray}
$\phi_{{\bf X}_i}$ and $\chi_i$ are spatial and spin functions, and 
$\tau_i$ is the isospin
function fixed to be up (proton) or down (neutron). 
Accordingly, an AMD wave function
is expressed by a set of variational parameters, ${\bf Z}\equiv 
\{{\bf X}_1,{\bf X}_2,\ldots, {\bf X}_A,\xi_1,\xi_2,\ldots,\xi_A \}$,
indicating single-nucleon Gaussian centroids and spin orientations for all
nucleons. 

These parameters are determined by the energy variation 
after spin-parity projection to obtain optimized AMD wave functions for $J^\pi$ states.
Namely, in the AMD+VAP method, 
the parameters ${\bf X}_i$ and $\xi_{i}$($i=1\sim A$) for the lowest $J^\pi$ state 
are determined so as to minimize the energy expectation value of the Hamiltonian,
$\langle \Phi|H|\Phi\rangle/\langle \Phi|\Phi\rangle$,
with respect to the spin-parity eigen wave function projected 
from an AMD wave function; $\Phi=P^{J\pi}_{MK}\Phi_{\rm AMD}({\bf Z})$.
Here, $P^{J\pi}_{MK}$ is the spin-parity projection operator.

In the present calculation, we choose the width parameter $\nu$ for single-nucleon Gaussian wave packets
to minimize energies of stable nuclei ($^9$Be, $^{11}$B, and $^{12}$C) and use 
the fixed $\nu$ value in each series of isotopes. The adopted $\nu$ values are $\nu=0.20$ fm$^{-2}$
for Be isotopes, and $\nu=0.19$ fm$^{-2}$ for B and C isotopes. 
The fixing $\nu$ parameter may not be appropriate 
to describe details of neutron distribution in very neutron-rich nuclei. However,  
since our main concern in the present study is systematics of proton distribution, 
we fix the parameter to remove a possible artifact in proton radii caused by 
the change of $\nu$. If the size of cluster cores in neutron-rich nuclei does not change from that in 
stable nuclei,  the fixing $\nu$ can be a reasonable assumption. For more detailed study, 
different width parameters for protons and neutrons
or independent widths for all nucleons should be adopted as done in the method of 
fermionic molecular dynamics (FMD) \cite{Feldmeier:1994he,Neff:2002nu}
and an extended version of AMD \cite{furutachi09}.   

In the AMD framework, the existence of clusters is not assumed {\it a priori}, but Gaussian centroids of
all single-nucleon wave packets are independently treated. 
Nevertheless, if the system favors a specific cluster structure 
such the structure is automatically obtained by the energy variation 
because the AMD model space contains wave functions for various cluster structures. 

We comment here that, in the simple AMD used in Refs.~\cite{KanadaEnyo:1995tb,KanadaEnyo:1995ir}, 
the energy variation was performed not after but before the spin projection 
(the variation before projection:VBP) for 
the AMD wave function with fixed single-nucleon intrinsic spins.
In the present study, an advanced method, the AMD+VAP, in which the VAP is performed for 
the AMD wave function with flexible intrinsic spins, is adopted. 
The AMD+VAP method better  describes structures of the ground and excited states of 
light nuclei and also useful to investigate details of the structure change between shell-model-like 
states and cluster states than the simple AMD.

Note that the AMD wave function is similar to the wave function used in FMD calculations \cite{Neff:2002nu},
though some differences exist in the width parameter and 
variational procedure, as well as adopted effective interaction. 

\section{Results}\label{sec:results} 

\subsection{effective interactions}
In the present calculation of Be and B isotopes, 
we used the same effective nuclear interaction as that used for 
$^{11}$Be and $^{12}$Be in previous studies \cite{ENYObe11,KanadaEn'yo:2003ue}.
It is the MV1 force \cite{MV1} for the central force 
supplemented by a two-body spin-orbit force with the two-range Gaussian form 
same as that in the G3RS force \cite{LS}.
The Coulomb force is approximated using a seven-range
Gaussian form. 
Namely, we use the interaction parameters, 
$m=0.65$, $b=0$, and $h=0$, for the Majorana, Bartlett, 
and Heisenberg terms of the central force, and the 
strengths $u_{I}=-u_{II}=3700$ MeV of the spin-orbit force
in the calculation of Be and B isotopes.
The breaking of the $N=8$ magicity
in $^{11}$Be and $^{12}$Be is successfully described with this set of interaction parameters 
as discussed in the previous studies \cite{ENYObe11,KanadaEn'yo:2003ue}.
For C isotopes, we use $m=0.62$ and $b=h=0$  for the central force, 
which is the parametrization same as that used for $^{12}$C in the previous
AMD+VAP calculation \cite{KanadaEn'yo:1998rf,KanadaEn'yo:2006ze}. In the present calculation of C isotopes, 
we tune the spin-orbit force strength and use 
$u_{I}=-u_{II}=2600$ MeV so as to reproduce the experimental excitation energies of the $2^+_1$ states
in C isotopes.

\subsection{Experimental data of rms proton and matter radii}
In the comparison of the calculated proton radii with the experimental data, 
we reduce the rms proton radii ($r_p$) from the rms charge radii ($r_c$) determined by  
isotope shift measurements as,  
\begin{equation}
r_p=\sqrt{r^2_c-R^2_p},
\end{equation}
where $R_p=0.8$ fm is the rms charge radii of an isolate proton.
The experimental data of the charge radii for Be isotopes,  $^{11}$B, and $^{12,14}$C 
are taken from Refs.~\cite{Nortershauser:2008vp,Krieger:2012jx,angeli04}.

In the experimental studies of the charge changing interaction cross section ($\sigma_{\rm cc}$), 
the proton radii have been deduced from a Glauber model analysis of the $\sigma_{\rm cc}$. 
We label thus deduced proton radii as $r_{\rm cc;G}$ in the present paper. 
The $r_{\rm cc;G}$ of neutron-rich B isotopes have been deduced from the  $\sigma_{\rm cc}$ at $\sim$900 MeV/u in
Ref.~\cite{estrade14}, and those of neutron-rich C isotopes have been deduced 
from the $\sigma_{\rm cc}$ at $\sim$300 MeV/u in Ref.\cite{Yamaguchi:2011zz}.

We also perform a rough evaluation of the proton radii of B and C isotopes 
from the experimental data of the $\sigma_{\rm cc}$ at $\sim$900 MeV/u on the C target in Ref.~\cite{chulkov00} using a following simple ansatz, 
\begin{equation}\label{eq:evaluated-rp}
\sigma_{\rm cc}=F \pi(r_p+r_{m,^{12}{\rm C}})^2,
\end{equation}
where $r_{m,^{12}{\rm C}}$ is the rms matter radius of the target nucleus, $^{12}$C, and $F$ is the normalization factor
for this beam energy.
We assume $r_{m,^{12}{\rm C}}$ equals to the proton radius $r_p$ of $^{12}$C, which is experimentally 
known from the charge radius, and determine the factor $F$ by the $\sigma_{\rm cc}$ for $^{12}$C beam
in the same experiment..
Using the common factor $F$ determined by the inputs of $r_p$ and  $\sigma_{\rm cc}$ for $^{12}$C, 
we evaluate the proton radii for B and C isotopes from the $\sigma_{\rm cc}$ in Ref.~\cite{chulkov00}. We call thus evaluated proton radii 
with the simple ansatz of Eq.~\ref{eq:evaluated-rp} as $r_{\rm cc:S}$.
Since there are many available data of the $\sigma_{\rm cc}$ for 
various neutron-rich isotopes in Ref.~\cite{chulkov00},  this evaluation is helpful to see 
the $N$ dependence of proton radii up to $N=14$ in B and C isotopes.

As for the rms matter radii, the radii $r_{\rm I}$ were deduced from the interaction cross section $\sigma_{\rm I}$ using 
the Glauber analysis \cite{ozawa2001}. Consistency of the matter radii determined by the Glauber analysis at various
beam energies has been checked (see Ref.~\cite{ozawa2001} and references therein).

\subsection{Be and B isotopes}
We perform the AMD+VAP calculation for the ground states of Be and B isotopes.
For $^{12}$Be, in which two $0^+$ states degenerate in the low-energy region, 
we also calculate the $0^+_2$ state by the VAP with respect to
the orthogonal component to the $0^+_1$ state, and superpose the obtained 
two AMD wave functions for the $0^+_1$ and $0^+_2$ states 
to take into account mixing of the configurations. 

\begin{figure}[tb]
\begin{center}
	\includegraphics[width=5.5cm]{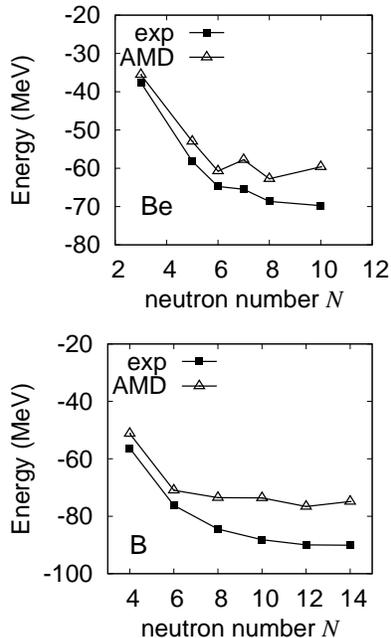} 	
\end{center}
\vspace{0.5cm}
  \caption{Binding energy of Be and B isotopes. 
The theoretical values are calculated with the AMD+VAP
using MV1($m=0.65$)+LS($u_{I}=-u_{II}=3700$ MeV) force.
$\nu=0.20$ fm$^{-2}$ and 0.19 fm $^{-2}$ are
used for Be and B isotopes, respectively. 
\label{fig:be-ene}.}
\end{figure}

Figure \ref{fig:be-ene} shows the binding energy of Be and B isotopes. 
The reproduction of the experimental biding energy in the present calculation is 
not perfect because of the limitation of the effective interaction. The reproduction can be 
improved by fine tuning of the interaction parameters or by introducing mass dependent interaction 
parameters. 
However, in the present study, we use the same parameters as the previous studies, 
which can describe the breaking of neutron magicity, to discuss the 
structure change long the isotopes, focusing on structure of protons. 

\begin{figure}[tb]
\begin{center}
	\includegraphics[width=5.5cm]{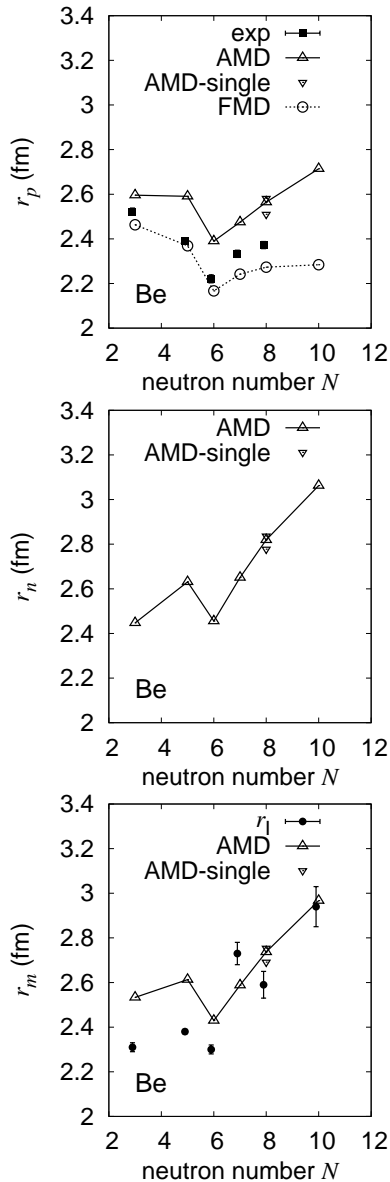} 	
\end{center}
\vspace{0.5cm}
  \caption{
\label{fig:be-radii}
Proton radii, neutron radii, and matter radii calculated with the AMD+VAP. 
For $^{12}$Be, the radius calculated with 
the single AMD wave function for each of the $0^+_1$ and $0^+_2$ states 
before the superposition is also shown (AMD-single). The radii of AMD-single for the $0^+_1$ are 
almost equal to those for the ground state after the superposition.
The experimental proton radii are those reduced from the experimental charge radii 
\cite{Nortershauser:2008vp,Krieger:2012jx,angeli04}. The experimental matter radii ($r_{\rm I}$) deduced from 
the interaction cross section \cite{ozawa2001} are also shown.
}
\end{figure}

Figure \ref{fig:be-radii} shows the rms radii of 
proton, neutron, and matter distributions of Be isotopes.  
For $^{12}$Be, we show radii calculated after and before the 
superposition of two AMD wave functions for $0^+_{1,2}$ obtained by the VAP. 
The proton radius is relatively large in $^7$Be and also in $^9$Be because of the 
remarkable cluster structure.
As the neutron number $N$ increases, the proton radius becomes the smallest in $^{10}$Be at $N=6$
and it increases in $^{11}$Be and $^{12}$Be, which have dominantly the intruder neutron configuration, and
becomes larger in  $^{14}$Be. The increase of the proton radii in the $N\ge 6$ region reflects 
the development of cluster structure.

The $N$ dependence of the  proton radius is consistent with the experimental 
data reduced from the charge radii determined by isotope shift measurements.
The trend of the $N$ dependence of the present result is also similar to the 
FMD predictions in the $N\le 8$ region \cite{Krieger:2012jx}. For $^{14}$Be, the present calculation 
predicts an increase of the proton radius because of the further development of
the cluster structure and deformation, 
whereas the FMD calculation does not show such an increase in $^{14}$Be. 

In the present result, the neutron radius grows more rapidly in the $N\ge 6$ region
as $N$ increases than the proton radius.
The $N$ dependence of the matter radius, which is mainly determined by that of the neutron radius, 
is consistent with the experimental matter radii $r_{\rm I}$ deduced from the interaction cross section \cite{ozawa2001}
except for a jump at $N=7$ in the experimental data. 
The extremely large matter radius in $^{11}$Be is caused by the neutron-halo structure, which 
is not described well in the present calculation because the wave function is limited to 
a Gaussian form and is not enough to describe the long tail of the halo neutron in the framework of the AMD+VAP.

\begin{figure}[tb]
\begin{center}
	\includegraphics[width=5.5cm]{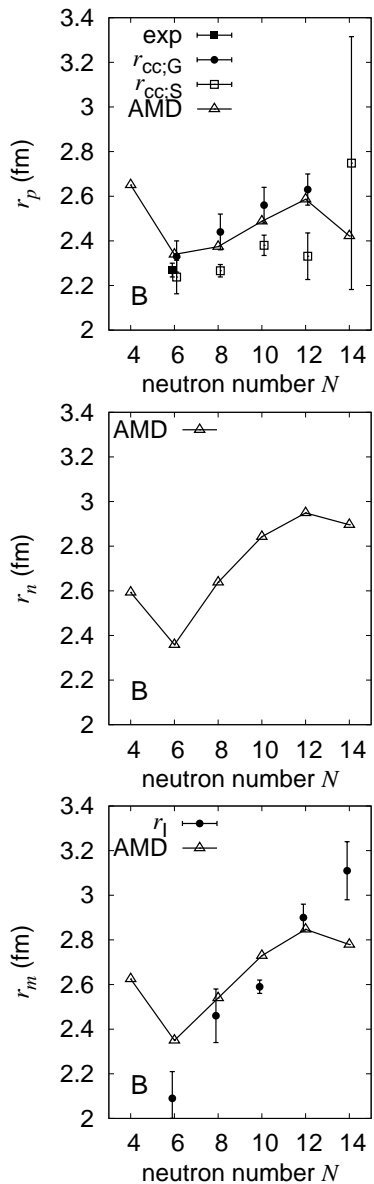} 	
\end{center}
\vspace{0.5cm}
  \caption{Proton radii, neutron radii, and matter radii calculated with the AMD+VAP. 
The experimental proton radius for $^{11}$B is reduced from the experimental charge radius 
\cite{angeli04}. The proton radii $r_{\rm cc;G}$ deduced from the charge changing interaction cross section 
$\sigma_{\rm cc}$ by the Glauber analysis in Ref.~\cite{estrade14}, and 
the proton radii $r_{\rm cc:S}$ evaluated from $\sigma_{\rm cc}$ in Ref.~\cite{chulkov00} 
using Eq.~\ref{eq:evaluated-rp} are also shown. 
The experimental matter radii ($r_{\rm I}$) are those deduced from 
the interaction cross section \cite{ozawa2001}.
\label{fig:b-radii}}
\end{figure}

\begin{table}[ht]
\caption{\label{tab:b-moment}
Electric quadrupole moments and magnetic dipole moments of B isotopes.
Theoretical values are calculated with the AMD+VAP. The experimental data are taken from 
Refs.~\cite{nucldata,Okuno1995,Ueno:1995hp,Izumi:1995mm,Ogawa:2003gp}. }
\begin{center}
\begin{tabular}{ccccc}
\hline
 &  \multicolumn{2}{c}{AMD} &  \multicolumn{2}{c}{exp.} \\
 & $\mu$($\mu_N$) & $Q$ (mb) & $\mu$($\mu_N$) & $Q$ (mb) \\
$^{9}$B  &2.65 &	6.54 &		 & 		\\	
$^{11}$B &2.79 &	3.90 &		2.689	&4.065(0.026)\\
$^{13}$B &2.97 &	3.65 &		3.178	&3.693(0.11)\\
$^{15}$B &2.64 &	4.15 &		2.650(0.013) &	3.80(0.10)\\
$^{17}$B &2.62 &	4.90 &		2.545(0.02)	& 3.86(0.15)  \\
$^{19}$B &2.75 &	3.79 &			&		\\
\hline
\end{tabular}
\end{center}
\end{table}

Figure \ref{fig:b-radii} shows the rms radii for B isotopes. As $N$ increases, 
the calculated proton radius becomes smallest at $N=6$ and increases 
in the $6\le N \le 12$ region in consequence of the developed cluster structure 
in the deformed neutron structure. 
The proton radius decreases from $N=12$ to $N=14$ because of the weakening of the
cluster structure in $^{19}$B. Note that the weakening of the cluster structure in 
$^{19}$B has not been obtained in the previous study in Ref.~\cite{KanadaEnyo:1995ir}, in which 
the adopted spin-orbit force was too weak to describe the shape coexistence in 
$N=14$ isotones \cite{KanadaEn'yo:2004cv}. 

In B isotopes, the charge radius is experimentally known only for $^{11}$B. 
We show,  in Fig.~\ref{fig:b-radii},  the experimental data of proton radii $r_{\rm cc;G}$ 
deduced from the charge changing interaction cross section
$\sigma_{\rm cc}$ by the Glauber analysis reported in Ref.~\cite{estrade14}. We also show 
the proton radii $r_{\rm cc;S}$ evaluated from $\sigma_{\rm cc}$ in Ref.~\cite{chulkov00} 
using Eq.~\ref{eq:evaluated-rp}.
The $N$ dependence of $r_{\rm cc;S}$ is consistent with that of $r_{\rm cc;G}$
for $^{11}$B, $^{13}$B, and $^{15}$B, but it is different at $N=12$ for $^{17}$B. 
The difference at $N=12$, in principle, comes from the discrepancy of the $\sigma_{\rm cc}$
between two experiments in Ref.~\cite{chulkov00} and Ref.~\cite{estrade14}.
The present calculation with the AMD+VAP shows the $N$ dependence consistent with $r_{\rm cc;G}$ deduced
from $\sigma_{\rm cc}$ in Ref.~\cite{estrade14}.
 
The neutron and matter radii show the $N$ dependence similar to each other. 
They show a kink at $N=6$ and the increasing behavior in the $6\le N \le 12$ region.
The experimental matter radii $r_{\rm I}$ deduced from the interaction cross section show a monotonic increase of matter radii
in the $6\le N \le 14$ region and are consistent with the present result except for $^{19}$B.
The present calculation probably underestimates the large neutron radius of $^{19}$B
caused by a neutron halo structure.

The present calculation predicts 
the kink at $N=6$ in the $N$ dependences of proton, neutron, and matter radii, which 
is consistent with the experimental 
proton and matter radii. It is interesting that the kink exists not at the $N=8$ magic number but at the $N=6$
in B isotopes.

In order to discuss the $N$ dependence of the proton radius around $N=12$ in more details, 
we also investigate moments of the $J^\pi=3/2^-$ ground states of B isotopes. 
Table \ref{tab:b-moment} shows the calculated electric quadrupole moments ($Q$) and magnetic moments ($\mu$) 
with the experimental data.  It is found that the present calculation reasonably reproduces the $Q$ moments of 
$^{11}$B, $^{13}$B, and $^{15}$B, but it overestimates the $Q$ moment of $^{17}$B. 
Since the experimental $\mu$ moment is smallest in $^{17}$B, 
it is likely that the contribution of the proton orbital angular momentum to 
the total spin $3/2^-$ is somewhat quenched in the realistic ground state of $^{17}$B, 
which usually reduces the $Q$ moment.
Another possibility is the weakening of the cluster structure in $^{17}$B, which 
reduces both the $Q$ moment and $r_p$. In the present calculation, 
no quenching of proton orbital angular momentum contribution nor the weakening of cluster structure is 
obtained in $^{17}$B.
A more precise measurement of proton radii of $^{17}$B is required. 

\subsection{C isotopes}

\begin{figure}[tb]
\begin{center}
	\includegraphics[width=5.5cm]{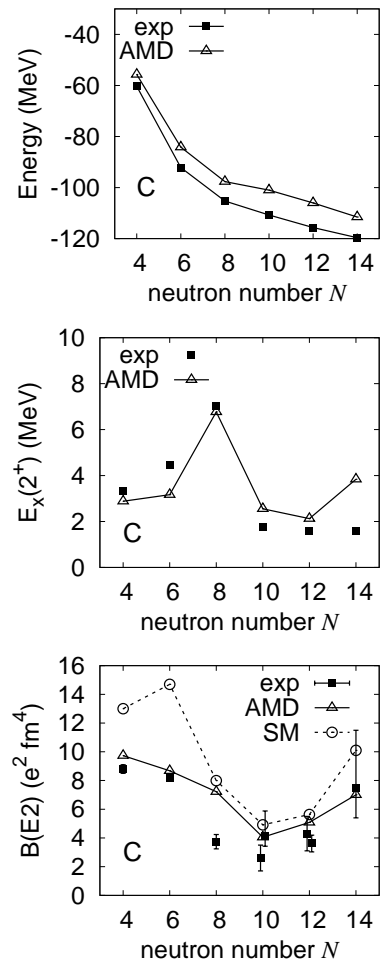} 	
\end{center}
\vspace{0.5cm}
  \caption{Binding energy, $2^+$ excitation energy, and 
$E2$ transition strength of C isotopes. 
The theoretical values are calculated with the AMD+VAP ($\nu=0.19$ fm$^{-2}$) 
using MV1($m=0.62$)+LS($u_{I}=-u_{II}=2600$ MeV) force.
The experimental data are taken from Refs.~\cite{nucldata,McCutchan:2012tw,Wiedeking:2008zzb,Ong:2007jb,Voss:2012zz,Petri:2011zz}.
Theoretical values for $B(E2)$ of the shell model calculation \cite{Sagawa:2004ut} are also shown.
\label{fig:c-ene-be2}}
\end{figure}

The $0^+_1$ and $2^+_1$ states of C isotopes are calculated with the 
AMD+VAP. 
Figure \ref{fig:c-ene-be2} shows the binding energy, the $2^+_1$ excitation
energy, and $B(E2;2^+_1\rightarrow 0^+)$ of C isotopes. 
The present calculation reasonably reproduces the experimental data
except for $E_x(2^+_1)$ in $^{20}$C and the $B(E2)$ value in $^{14}$C, which are 
overestimated by about a factor two. 

\begin{figure}[tb]
\begin{center}
	\includegraphics[width=5.5cm]{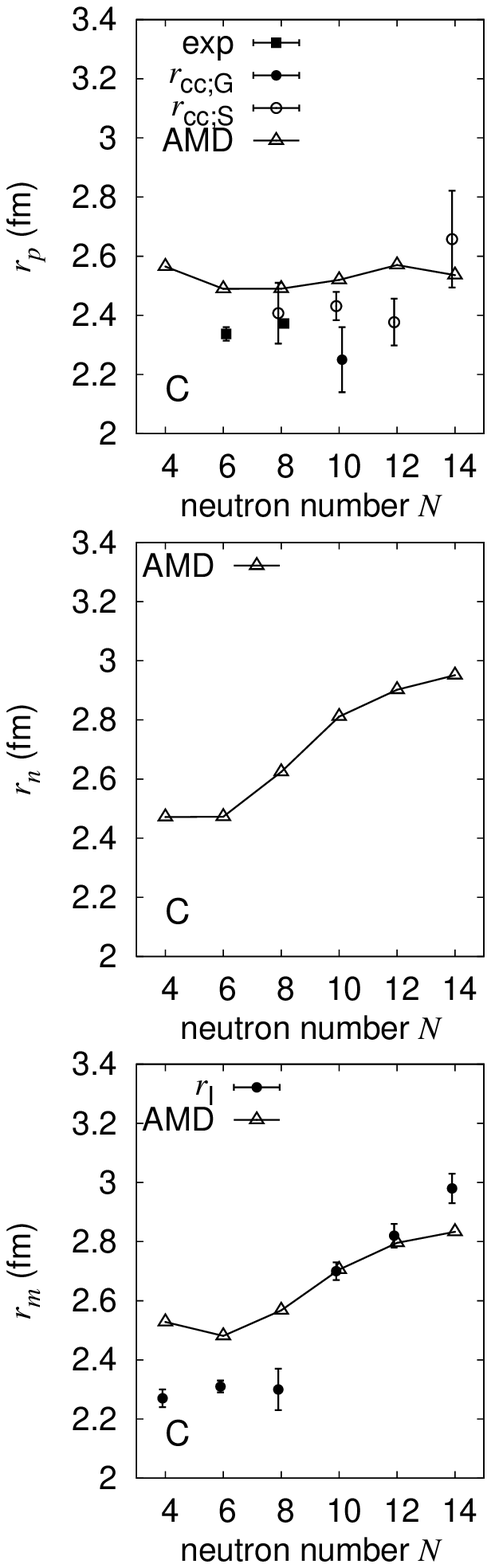} 	
\end{center}
\vspace{0.5cm}
  \caption{
Proton radii, neutron radii, and matter radii calculated with the AMD+VAP. 
The experimental proton radii for $^{12,14}$C are reduced from the experimental charge radius 
\cite{angeli04}. The proton radius $r_{\rm cc;G}$ of $^{16}$C deduced from the $\sigma_{\rm cc}$ by the Glauber analysis
in Ref.~\cite{Yamaguchi:2011zz}, and 
the proton radii $r_{\rm cc;S}$ evaluated 
from the $\sigma_{\rm cc}$ in Ref.~\cite{chulkov00} using 
Eq.~\ref{eq:evaluated-rp} are also shown. 
The experimental matter radii ($r_{\rm I}$) are those deduced from 
the interaction cross section \cite{ozawa2001}.
\label{fig:c-radii}}
\end{figure}

Figure \ref{fig:c-radii} shows the rms proton, neutron, and matter radii 
of C isotopes. Even though the neutron and matter radii 
increase in the $N\ge 6$ region as $N$ increases, 
the proton radius is almost unchanged. The weak $N$ dependence of the proton radius
indicates the insensitivity of the proton distribution to the neutron 
structure. This is contrast to the cases of Be and B isotopes having the rather strong
$N$ dependence of proton radii.
The $N$ dependence of the matter radius in the present result 
is consistent with the experimental $r_{\rm I}$ deduced from the interaction cross section.
The proton radii $r_{\rm cc;S}$ evaluated from the experimental data of the 
$\sigma_{\rm cc}$ at approximately 900 MeV/u show a weak $N$ dependence in the 
$8\le N \le 12$ region and seem to be consistent with the present prediction. 
There exists an experimental data of $r_{\rm cc;G}$ for $^{16}$C deduced from the $\sigma_{\rm cc}$ at approximately 300 MeV
by the Glauber analysis ~\cite{Yamaguchi:2011zz} seems to somewhat deviate from other data. 

\section{Discussions}\label{sec:discussion}
In this section, we describe the intrinsic structure change with the increase of the neutron number 
in each series of isotopes and discuss its effect to the $N$ 
dependence of proton radii.
 
\begin{figure}[tb]
\begin{center}
	\includegraphics[width=15.0cm]{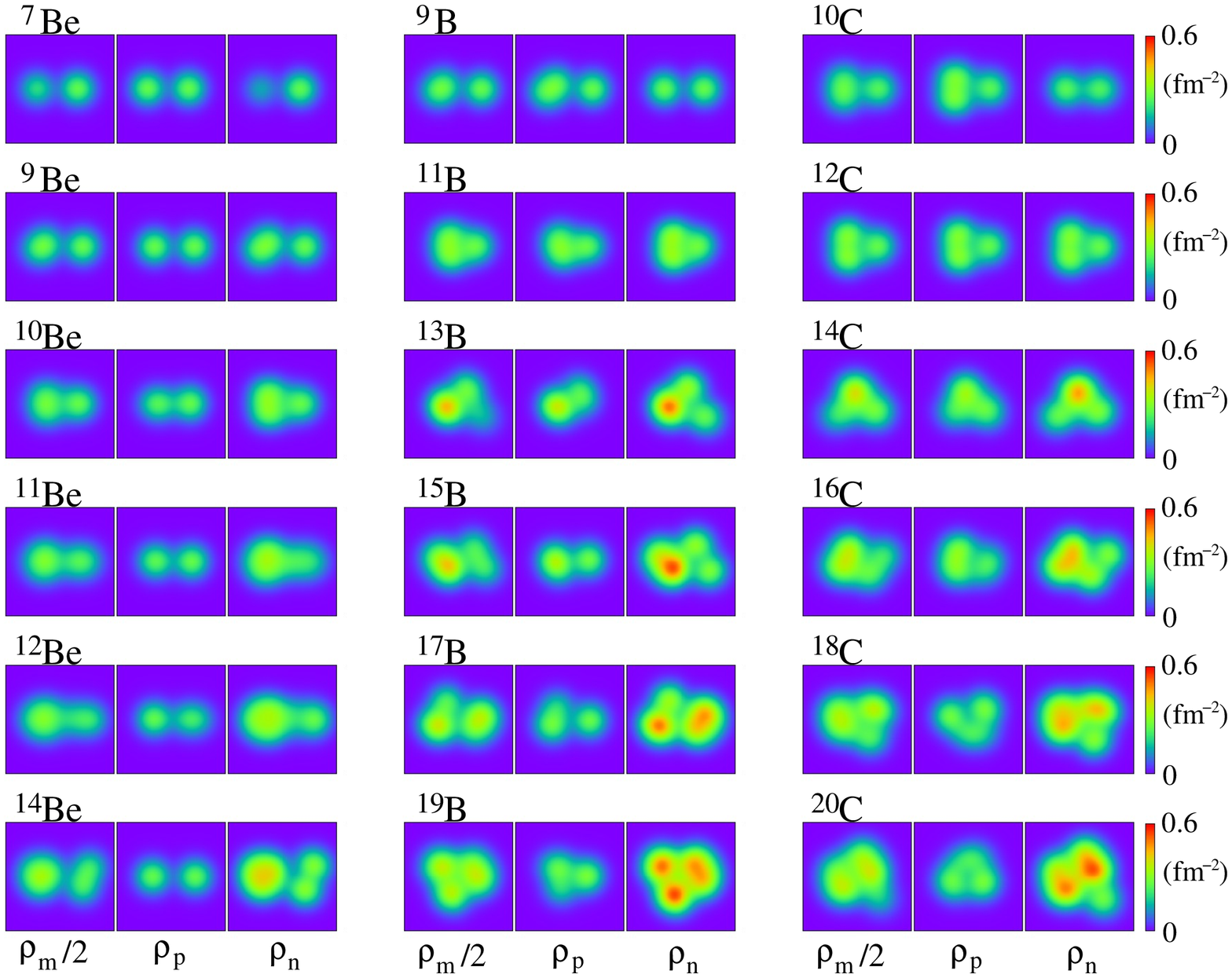} 	
\end{center}
\vspace{0.5cm}
  \caption{(Color online) 
Distributions of proton, neutron, and matter densities calculated with the AMD+VAP.
The densities of intrinsic states are integrated with respect to the $z$ axis and
plotted on $x$-$y$ plane. Here, the axes of the intrinsic frame are chosen so as to be
$\langle x^2\rangle \ge \langle y^2\rangle \ge \langle z^2\rangle$.  
\label{fig:bebc-dense}}
\end{figure}

Figure \ref{fig:bebc-dense} shows the distributions of proton, neutron,
and matter densities of Be, B, and C isotopes
obtained by the AMD+VAP. The density distributions of intrinsic states before
the spin and parity projections are displayed.
In all series of Be, B, and C isotopes, 
the intrinsic neutron structures change rapidly with the increase of $N$. 

\begin{figure}[tb]
\begin{center}
	\includegraphics[width=5.5cm]{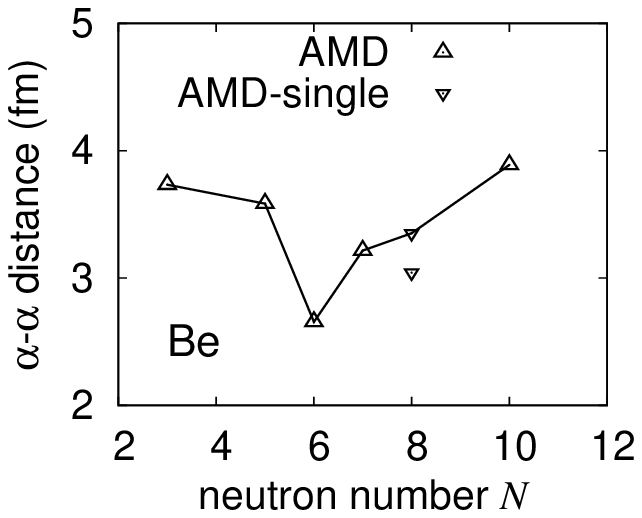} 	
\end{center}
\vspace{0.5cm}
  \caption{$\alpha$-$\alpha$ distance in Be isotopes. 
\label{fig:be-raa}}
\end{figure}

In Be isotopes, the $2\alpha$ cluster core structure is formed as 
shown in the dumbbell shape in the proton density. 
Following the development of the prolate neutron deformation,
the cluster structure in Be isotopes
is enhanced in the $7\le N \le 10$ region, resulting in the increase of the 
proton radius in this region.
Figure~\ref{fig:be-raa} shows the $\alpha$-$\alpha$ distance measured by Gaussian centroids for four protons as 
$|{\bf X}_1+{\bf X}_2-{\bf X}_3+{\bf X}_4|/2\sqrt{\nu}$, which indicates a degree of the $2\alpha$ cluster development in Be isotopes.
The $\alpha$-$\alpha$ distance describes 
the $N$ dependence of the proton radius in Be isotopes.  

In B isotopes, the neutron density is most compact at $N=6$ for $^{11}$B because of the 
$p_{3/2}$ sub-shell closure feature. Also the proton structure in $^{11}$B is compact 
and shows no cluster structure, whereas,
in neutron-rich B isotopes with $N\ge 8$, 
the two-center cluster structure develops as shown in the proton 
distribution. The development of the cluster structure is remarkable 
at $N=10$ and $N=12$ for $^{15}$B and $^{17}$Be resulting in the 
enhanced proton radii of these nuclei, whereas it slightly weakens 
in $^{19}$Be. 

In C isotopes, the proton density always stays in a compact region in neutron-rich C
with $N\ge 8$ even though the neutron structure rapidly changes with the increase of $N$. 
It indicates the robustness of the proton structure of $Z=6$ system 
in neutron-rich C isotopes, in which protons are deeply bound.
The stable proton structure is reflected in the weak $N$ dependence of 
the proton radius. 

As discussed above, in neutron-rich Be and B isotopes, which have two-center cluster structures, 
the proton structure changes sensitively to the neutron structure change.
In contrast, in C isotopes, the proton structure is insensitive to the neutron structure and has the 
weak $N$ dependence. The sensitivity of the proton structure to the neutron structure is essential 
in the $N$ dependence of the proton radius. 
Development and weakening of the two-center cluster structures in Be and B isotopes 
play an important role in the change of proton radii with the $N$ increase.

\begin{figure}[tb]
\begin{center}
	\includegraphics[width=6cm]{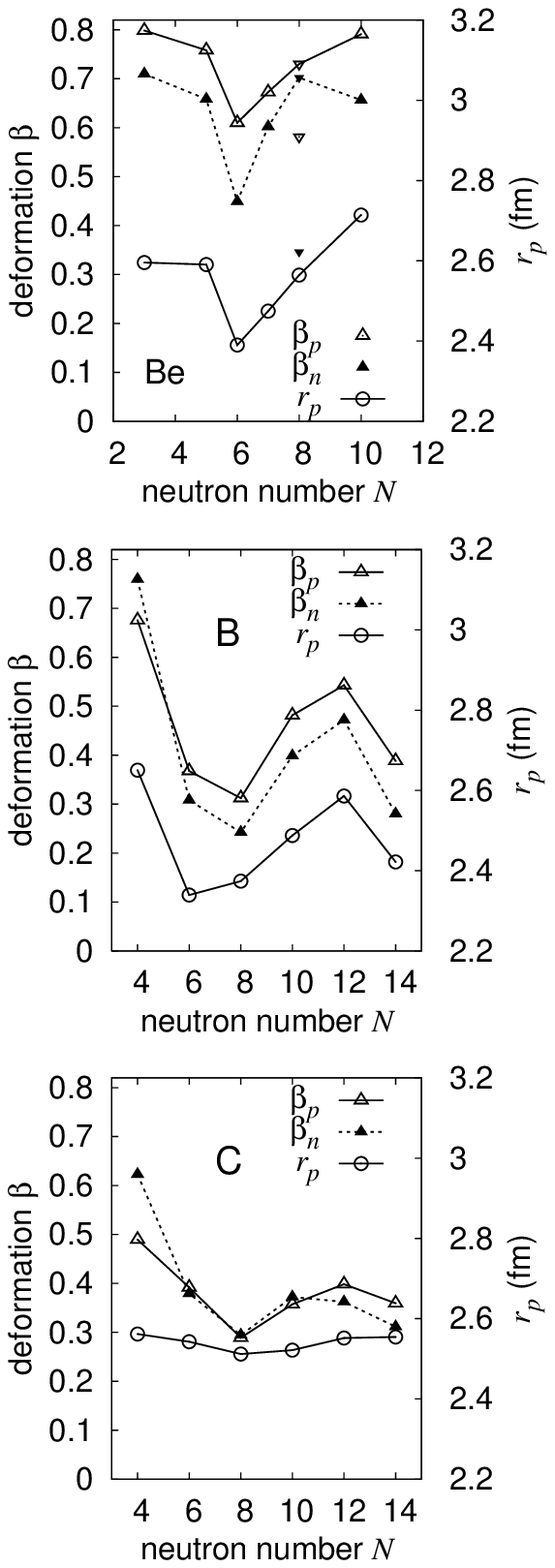} 	
\end{center}
\vspace{0.5cm}
  \caption{Deformation parameter $\beta$ for proton, neutron, and matter densities, and proton radii of 
Be, B, and C isotopes calculated with the AMD+VAP.
\label{fig:be-defo-rp}}
\end{figure}

To see how the neutron structure change affects the $N$ dependence of proton radii through the
proton structure change, we show, in Fig.~\ref{fig:be-defo-rp}, the $N$ dependence of the deformation 
parameters $\beta_p$ and $\beta_n$ for proton  
and neutron densities, respectively, in the intrinsic wave functions compared with the $N$ dependence 
of proton radii in Be, B, and C isotopes.
Here, the definition of $\beta$ is that defined in Ref.~\cite{KanadaEn'yo:1996hi}. 

In Be isotopes, the change of the proton deformation correlates with the neutron deformation 
except for $^{14}$Be at $N=10$. In  $^{14}$Be, the neutron deformation is not as large as 
that in $^{12}$Be, but the wide distribution of the neutron density stretches 
the two-center proton density, resulting in the larger proton deformation than that in $^{12}$Be. 
The proton deformation just describes the $N$ dependence of the proton radius in Be isotopes.

Also in B isotopes, the change of proton deformation strongly correlates with the neutron deformation. 
The $N$ dependence of the proton deformation is consistent with that of the proton radius in the neutron-rich 
$N\ge 8$ region, in which B isotopes have the two-center cluster structure as mentioned previously. However, 
in the region from $N=6$ to $N=8$, the $N$ dependence of the proton radius is opposite 
to that of the proton deformation. Namely, the proton radius slightly increases from $^{11}$B to $^{13}$B
even though the deformation becomes small at the neutron magic number $N=8$.  
As discussed in the previous section, since
$^{11}$B has the smallest neutron radius and no cluster structure, it has the smallest proton radius
in B isotopes. 

In C isotopes, the change of $\beta_p$ is consistent with $\beta_n$. 
Note that the consistency between $\beta_p$ and $\beta_n$ does not necessarily mean
the consistency in the shapes between proton and neutron density distributions but the $\gamma$ parameters
for proton and neutron distributions are different from each other in some C isotopes.
The $N$ dependences of proton and neutron deformations in C isotopes 
are weaker than those in Be and B isotopes. Moreover, the change of proton deformation makes only 
the small change of the proton radii. This situation of neutron-rich C isotopes having no cluster structure
is different from the cases of neutron-rich Be and B isotopes having two-center cluster structures, in which 
the proton radius correlates with the proton deformation.   
As a result, proton radii in C isotopes are insensitive to 
the neutron structure change and do not depend so much on the neutron number.
The weak $N$ dependence of the proton radius in C isotopes is considered to 
originate in stable oblate proton deformation and the non-cluster structure. 

In the systematic analysis of the structure change and its effect on proton radii in Be, B, and C isotopes,  
we can reach the more general picture that, in light nuclei,  
the strong $N$ dependence of proton radii is found in the isotopes that have prolate deformations in
both proton and neutron densities. In neutron-rich Be and B isotopes, 
the prolate proton deformation is caused by the development
of two-center cluster structure.
Since the cluster structure can be easily stretched by the prolate neutron deformation, 
the central proton density becomes low and the proton radii can be enhanced.
In other words, the decrease of the central proton density 
in the developed cluster structure in neutron-rich nuclei is important in the 
sensitivity of proton radii to the structure change. Consequently, 
the $N$ dependence of proton radii can be a probe to observe development of cluster structure. .

\section{Summmary}\label{sec:summary} 

We investigated the $N$ dependence of proton radii of Be, B, and C isotopes. 
In the result of the AMD+VAP calculation for Be and B isotopes,  we found that 
the proton radius sensitively reflects the neutron structure change through the 
the development of cluster structure, in particular, in neutron-rich nuclei.
In contrast, the proton radius in C isotopes shows a weak $N$ dependence 
because of the stability of the proton structure in $Z=6$ nuclei.
We compared the $N$ dependence of the calculated 
proton radii with that of the experimental radii reduced from the 
charge radii measured by means of isotope shift and those deduced from 
the charge changing interaction cross section, and found that the present result is
consistent with the existing experimental data.

In the analysis of the structure change and its effect on proton radii in Be, B, and C isotopes, 
we found that the $N$ dependence of proton radii can be a probe to clarify enhancement 
and weakening of cluster structures.
In neutron-rich Be and B nuclei, the 
two-center cluster structure is enhanced in the prolately deformed neutron structure.
The $N$ dependence of proton radii reflects rather sensitively the 
cluster structure change, because the central proton density becomes low 
in consequence of  the stretching of the cluster structure.
Precise measurements of proton radii for B and C isotopes are required to confirm the cluster structure 
in neutron-rich B isotopes and the non-cluster structure in C isotopes..

\section*{Acknowledgments} 
The author would like to thank Prof.~Tanihata and Prof.~Kanungo for fruitful discussions.
She also thanks Prof.~Kimura for valuable comments. 
The computational calculations of this work were performed using the
supercomputer at YITP.
This work was supported by 
JSPS KAKENHI Grant Number 26400270.

\end{document}